# Unusual Low Temperature Reactivity of Water. The CH + H$_2$O Reaction as a Source of Interstellar Formaldehyde?

*Kevin M. Hickson*[1,2]\*, *Philippe Caubet*[1,2] *and Jean-Christophe Loison*[1,2]

[1]Université de Bordeaux, Institut des Sciences Moléculaires, Unité Mixte de Recherche 5255, F-33400 Talence, France.

[2]CNRS, Institut des Sciences Moléculaires, Unité Mixte de Recherche 5255, F-33400 Talence, France.

**Corresponding Author**

\* Correspondence to: km.hickson@ism.u-bordeaux1.fr

Water is an important reservoir species for oxygen in interstellar space and plays a key role in the physics of star formation through cooling by far-infrared emission. Whilst water vapour is present at high abundances in the outflows of protostars, its contribution to the chemical evolution of these regions is a minor one due to its limited low temperature reactivity in the gas-phase. Here, we performed kinetic experiments on the barrierless CH + H$_2$O reaction in a supersonic flow reactor down to 50 K. The measured rate increases rapidly below room temperature, confirming and extending the predictions of earlier statistical calculations. The open product channels for this reaction suggest that this process could be an important gas-phase route for formaldehyde formation in protostellar envelopes.

As one of the dominant reservoirs for interstellar oxygen, the water molecule plays a pivotal role in the chemistry and physics of the universe. In cold, dense molecular clouds, where atomic oxygen abundances are high, gas-phase water formation occurs through a series of ion-molecule reactions beginning with the O + $H_3^+$ reaction and ending with the dissociative recombination reaction of the $H_3O^+$ ion. Whilst gas-phase chemistry is still considered to be an important source of interstellar water, the traditional view of water formation mechanisms has changed in recent years. It is now clear from surface science experiments that water is formed efficiently on interstellar dust grains through the reactions of accreted hydrogen and oxygen atoms which are effectively promoted by surface catalysis.[1] As the dense cloud temperature (10 K) is significantly lower than the assumed ice sublimation temperature (90-100 K), most of the water is present as ice, so that observed gas-phase water abundances relative to $H_2$ are low ($10^{-9}$-$10^{-10}$).[2] As the molecular cloud evolves to higher densities, gravitational collapse leads to the formation of an embedded protostar whose luminosity heats the surrounding gas and dust to temperatures above 100 K. Under these conditions, water ice present in the grain mantles sublimates returning it to the gas-phase where it acts as a major coolant through emission from its excited rotational levels at far-infrared wavelengths, sustaining protostellar growth. In these regions, astrochemical models predict that $H_2O$ is destroyed primarily by photodissociation and through its reactions with cations such as $H^+$, $He^+$ and $H_3^+$. Nevertheless, these processes simply transform water into other temporary oxygen hydride reservoir species which mostly reform water.[3]

Whilst water has been shown to catalyze reactions between neutral species due to its ability to lower intrinsic barriers,[4] few reactions of the water molecule itself with neutral species are considered in astrochemical databases[5] given its low inherent reactivity below room temperature. Indeed, the only reaction of a ground-state neutral radical with water currently included is the F + $H_2O$ reaction. However, as this process only occurs through tunneling due to the presence of an activation barrier,[6] it will be relatively ineffective at low temperature. To date, the only ground-state neutral radical to demonstrate an apparent increasing reactivity with water with decreasing temperature from 700–300 K is the methylidyne radical, CH.[7,8] The CH radical in its X($^2\Pi_r$) ground state is highly reactive towards saturated and unsaturated molecules,[9] playing an important part in carbon chain formation through insertion into C-H and C-C bonds. On Titan, the largest of Saturn's moons and the only satellite in the solar system possessing a significant atmosphere, methylidyne is generated by the photodissociation of methane[10] with its production peak occurring at altitudes around 800 km. Interestingly, water is also deposited at altitudes around 750 km through the ablation of micrometeorites,[11] so that this reaction could influence oxygen chemistry in Titan's atmosphere. In the gaseous envelopes surrounding young stellar objects (YSOs), a wide range of simple hydride species including CH have been detected.[12] In these same sources, gas-phase water abundances relative to $H_2$ reach $10^{-4}$ [13] suggesting that the neutral chemistry of water could play an important role in astrochemistry, if two conditions are fulfilled. Firstly, given the low total densities typically found in these astrochemical environments, the CH + $H_2O$ system must be shown to result in real product formation rather than just third-body assisted complex or adduct stabilization. Secondly, the reaction must be shown to occur rapidly below room temperature.

Whilst the reaction of CH with $H_2O$ proceeds initially through a pre-reactive, HC⋯$OH_2$ complex,[14,15] recent statistical calculations[15] predict that the complex passes through a transition state (TS) lower in energy than the reagent asymptote to form the H$_2$COH adduct. The adduct should then evolve quickly to yield two of three energetically accessible sets of products,

CH + H$_2$O ⇆ HC···OH$_2$ → H$_2$COH → H + H$_2$CO    ΔH$_r$ = -241.8 kJ/mol    (R1a)

→ H$_2$ + HCO    ΔH$_r$ = -328.9 kJ/mol    (R1b)

with the energetic HCO product formed by channel (1b) falling apart to give H + CO, explaining the experimentally observed H atom yield of 100 % at 330 K.[15] Although this measurement confirms that product formation occurs, it does not allow us to differentiate between individual product channels. In this respect, experimental and theoretical studies of the evolution of the H$_2$COH intermediate produced by the O + CH$_3$ reaction[16] indicate that the H + H$_2$CO (formaldehyde) channel represents approximately 85% of the total products at room temperature with little predicted temperature dependence. Given the only slightly lower energy of the CH + H$_2$O reactants (-42 kJ/mol) with respect to the O + CH$_3$ ones and the large calculated exothermicities of product channels R1a and R1b, it is likely that product branching ratios for reaction R1 will resemble closely those obtained for the O + CH$_3$ reaction.[16]

The apparent rate of reaction R1 is governed by the competition between adduct formation and redissociation to reactants in a similar manner to other complex forming reactions.[17] Statistical calculations predict that complex redissociation is inhibited at low temperature,[15] so that reaction should be dominated by the long range attractive part of the intermolecular potential. If these predictions are correct, the rate constant should increase monotonically with decreasing temperature, eventually reaching the capture rate limiting value. Reaction R1 could thus represent a substantial source of gas-phase H$_2$CO in low temperature environments. Nevertheless, no experimental kinetic studies of reaction R1 have ever been performed below room temperature to validate this hypothesis.

To address this issue, and to assess it potential importance for astrochemistry, we investigated the kinetics of the CH + H$_2$O reaction from 296 K down to 50 K using a CRESU apparatus,[18] described briefly in the Experimental Methods. More detailed information can be found in the Supporting Information (SI) file.

Representative decays of the CH signal are shown in Figure 1 for experiments conducted at 50 K.

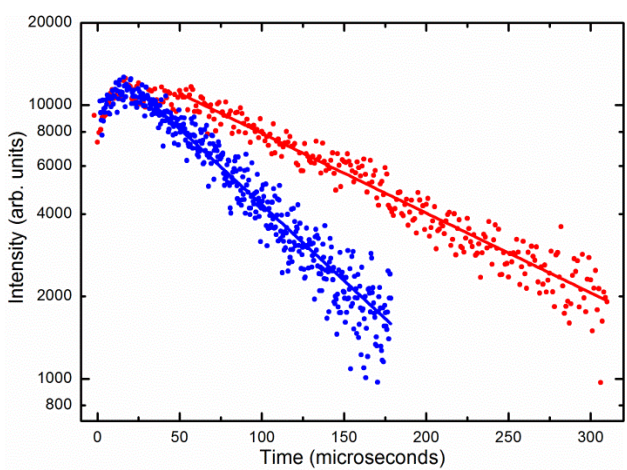

**Figure 1.** Typical decay profiles of the CH (OH*) chemiluminescence intensity as a function of time taken at 50 K. (●) In the absence of water vapour; (●) with [H$_2$O] = 3.3 × 10$^{13}$ molecule cm$^{-3}$.

The data were fitted using a simple exponential function to yield the pseudo-first-order rate constant, $k_1'$ from the time constant of the decay. Second-order rate constants, $k_1$, were determined from a weighted linear least-squares analysis of plots of the individual $k_1'$ values versus the H$_2$O concentration. Figure 2 shows the measured $k_1'$ values as a function of the H$_2$O concentration at 50 K and 296 K in argon with their associated statistical uncertainties.

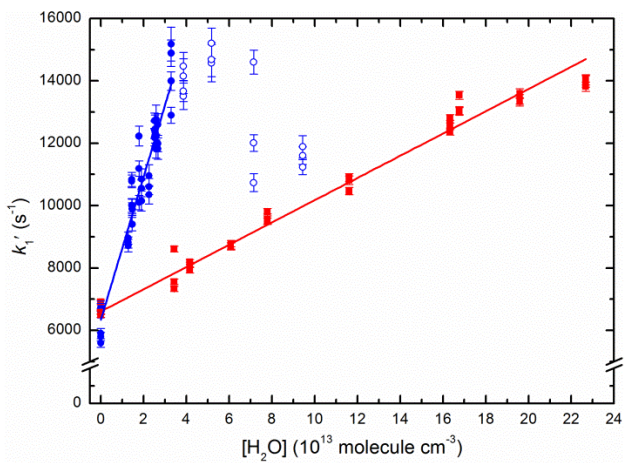

**Figure 2.** Plots of the $k_1'$ values at 296 K (■) and at 50 K (●) as a function of the H$_2$O concentration. Datapoints obtained at 50 K with H$_2$O concentrations above the threshold for cluster formation (○). The uncertainties are obtained from non-linear fits to decays similar to those presented in Figure 1. The solid lines represent fits using a $1/\sigma^2$ weighting function. $\sigma$ is the statistical uncertainty of individual $k_1'$ values.

The exploitable range of H$_2$O concentrations at 50 K in particular was severely limited by an apparent fall-off in the observed $k_1'$ values at water concentrations above a certain threshold. The blue open circles in Figure 2 show the results of experiments conducted at H$_2$O concentrations higher than this threshold. All experiments below room temperature were seen to be affected in this way, to a greater or lesser extent depending on the underlying physical characteristics of the flow. This effect was interpreted as the onset of cluster formation leading to a reduction in the free gas-phase water concentration. Only experiments conducted in the linear regime were used in the final analysis. The corresponding second-order plots obtained at all other temperatures are presented in Figures S1-S5 of the SI file.

The $k_1$ values obtained at each temperature are summarized in Table S1, with the respective carrier gas and H$_2$O concentration ranges used. Our results for the CH + H$_2$O reaction over the 50-296 K range are compared to previous work in Figure 3. The rate constant values previously obtained for the CH + H$_2$O reaction are listed in Table S2.

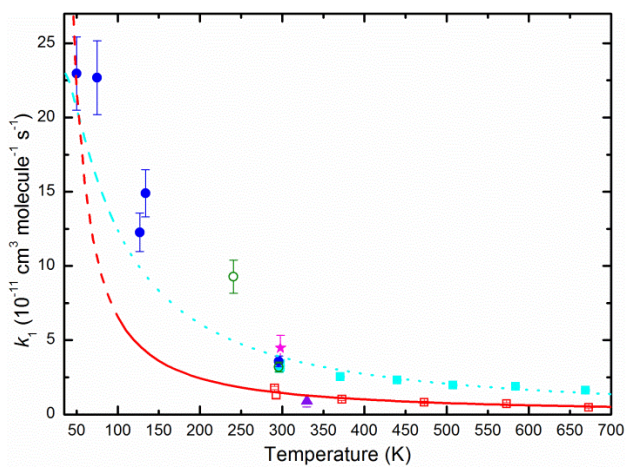

**Figure 3.** Rate constants for the CH + H$_2$O reaction as a function of temperature. (★) Bosnali and Perner;[19] (■) Zabarnick et al.;[8] (□) Blitz et al.;[7] (▲) Bergeat et al.;[15] (●) This work, Ar carrier gas; (○) This work, SF$_6$/N$_2$ carrier gas; (┄) Bergeat et al.,[15] TS energy -16.2 kJ/mol; (—) Bergeat et al.,[15] TS energy -7.2 kJ/mol. The error bars on the present values are the combined statistical and systematic uncertainties.

At 296 K our measured rate constants over the pressure range 2-20 Torr with Ar and SF$_6$/N$_2$ mixtures as the carrier gases are slightly lower than the value of Bosnali and Perner,[19] and slightly higher than the values of Blitz *et al.*[7] from 70-200 Torr. Excellent agreement is observed with the study of Zabarnick *et al.*[8] who obtained an average value of $k_1$(298 K) = 3.23 × 10$^{-11}$ cm$^3$ molecule$^{-1}$ s$^{-1}$ over a range of pressures from 20-300 Torr. Crucially, the large combined pressure range covered by these studies indicates that little or no stabilization of the prereactive HCOH$_2$ complex occurs at room temperature, in good agreement with calculations.[15] The present work confirms this result by employing an efficient collision partner at 296 K, SF$_6$, in the carrier gas mixture, which is characterized by an average energy loss per collision substantially larger than those of rare gases. Although the rate constants obtained by the earlier temperature dependent studies[7,8] differ by a factor of two, indicating possible systematic errors in the gas-phase water concentration measurement, both sets of rate constants increase with decreasing temperature, a trend that is supported by the statistical calculations[15] also displayed in Figure 3. Two different values of the TS energy for passage from the prereactive complex to the adduct were required in the calculations, -16.2 kJ/mol and -7.2 kJ/mol to furnish good agreement with the experiments of Zabarnick *et al.*[8] and Blitz *et al.*[7] respectively. The present study extends these measurements by demonstrating an order of magnitude increase of the rate constant between 296 K and 50 K; in good agreement with the calculated values employing a TS energy of -16.2 kJ/mol over the 296-100 K range. Indeed, the agreement between experiment and theory would be improved using a slightly lower TS energy. At 50 K, the measured rate constant approaches the thermal association rate value (calculated at 100 K) of $2.4 \times 10^{-10}$ cm$^3$ molecule$^{-1}$ s$^{-1}$,[15] indicating that virtually every collision results in reaction. Interestingly, using a threefold higher density Ar nozzle at 134 K (compared to the corresponding 127 K Ar nozzle) and a 241 K nozzle based on SF$_6$/N$_2$ mixtures, it has been possible to explore the influence of the carrier gas on the reaction rate. Although the measured 134 K rate constant is approximately 20% larger than the corresponding 127 K rate, the associated error bars overlap, providing only tentative evidence that quenching occurs. Nevertheless, the measured rate constant at 241 K of $(9.3 \pm 1.1) \times 10^{-11}$ cm$^3$ molecule$^{-1}$ s$^{-1}$ is two times higher than the calculated value of $5.0 \times 10^{-11}$ cm$^3$ molecule$^{-1}$ s$^{-1}$ using argon as the buffer gas with a TS energy of -16.2 kJ/mol.[15] Taken together, these observations suggest that redissociation of the HCOH$_2$ complex is partially inhibited by collisional relaxation at lower temperature leading to an enhancement of the overall rate, an effect which was not predicted by theory.[15]

This present case contrasts with that of the OH + acetone reaction which was also recently seen to present a substantial increase in reactivity at low temperature.[20] Here, passage from the prereactive complex to the adduct is hindered by the presence of a TS higher in energy than the reagent asymptote, so that the large reactivity increase is probably due to the efficient collisional stabilization of the prereactive complex as the temperature falls. Consequently, at the low densities prevalent in many interstellar environments, the complex will redissociate to reactants despite its relatively long lifetime, with only limited product formation by tunneling through the activation barrier.[21]

The question remains as to whether reaction R1 will play an important role in similar low temperature, low density astrochemical environments. At low temperature, the equilibrium between adduct formation and redissociation to reactants weighs heavily in favour of adduct formation given the low energy of the exit TS. Moreover, as the TSs for passage from the adduct to products are also low in energy,[15] the adduct should evolve to products even in the low pressure regime, acting as a source of gas-phase formaldehyde in regions with large gas-phase water abundances, if as predicted, the room temperature product branching ratios for channels R1a and R1b do not change at low temperature. Updated statistical calculations focusing on product formation from the $H_2COH$ adduct would be extremely useful in this respect, allowing low temperature branching ratios for channels R1a and R1b to be predicted. Furthermore, as the results of the present study place much tighter constraints on the potential values of the TS energy for $H_2COH$ formation, it will be possible to estimate the relative degrees of complex $HCOH_2$ and adduct $H_2COH$ stabilization as a function of pressure.

High abundances of formaldehyde molecules are observed in the envelopes of YSOs[22] and in photodissociation regions[23] which are attributed entirely to the evaporation of grain mantles through photo-desorption. Although grain surface reactions are likely to represent the major source of $H_2CO$, the $CH + H_2O$ reaction should also be considered in gas-grain models to evaluate its eventual contribution.

EXPERIMENTAL METHODS

The low temperature rate constants for reaction R1 were measured using the CRESU technique. This method is ideally suited to the study of low temperature reactions involving species which are nominally in the liquid phase at room temperature. As problems associated with condensation are absent as the cold supersonic flow is isolated from the reactor walls, supersaturated concentrations of such species are easy to attain. In contrast, cryogenically cooled apparatuses are entirely limited by the saturated vapour pressure of the liquid component of the flow below room temperature.

In these experiments, CH radicals were generated by pulsed laser photolysis and a chemiluminescent tracer method was used to follow the CH temporal evolution in the presence of an excess of water vapour. Water vapour itself was introduced into the reactor using a controlled liquid delivery system coupled with an online absorption cell to determine its *in-situ* concentration precisely. More detailed information regarding the methodology can be found in the accompanying SI file.

ASSOCIATED CONTENT

**Supporting Information Available**: Detailed experimental methodology. Supplemental figures S1-S5. Supplemental tables S1-S3. This material is available free of charge via the Internet at http://pubs.acs.org.

AUTHOR INFORMATION

**Corresponding Author**

*Email: km.hickson@ism.u-bordeaux1.fr

**Notes**

The authors declare no competing financial interests.

ACKNOWLEDGMENTS

This work was supported by the INSU-CNRS national programs PCMI and PNP, the Hubert Curien Program Galileo project number 28125ZM and the Observatoire Aquitain des Sciences de l'Univers.

# Supporting Information File for "The Unusual Low Temperature Reactivity of Water. The CH + $H_2O$ Reaction as a Source of Interstellar Formaldehyde?"

Kevin M. Hickson[1,2]*, Philippe Caubet[1,2] and Jean-Christophe Loison[1,2]

[1]Université de Bordeaux, Institut des Sciences Moléculaires, Unité Mixte de Recherche 5255, F-33400 Talence, France.

[2]CNRS, Institut des Sciences Moléculaires, Unité Mixte de Recherche 5255, F-33400 Talence, France.

**Experimental Methodology** The CRESU technique was used for the present study. As the method has been previously described,[25] only the experimental details specific to the current investigation will be outlined here. Two new Laval nozzles were employed during this study. The first of these allowed us to perform kinetic measurements at a specified temperature of 134 K in Argon but with a higher flow density than the corresponding 127 K nozzle, to enable us to check for pressure effects at low temperature. The second new nozzle based on a carrier gas mixture of $N_2$ and $SF_6$ allowed us to perform kinetic measurements at a temperature of 241 K, higher than ever previously attained with any CRESU apparatus. Given the high collisional efficiency of the $SF_6$ molecule, this nozzle also allowed us to check for potential stabilization effects closer to room temperature. The velocity and temperature of the supersonic flows calculated from separate measurements of the impact pressure using a Pitot tube and the stagnation pressure within the reservoir are summarized in Table S3 alongside other relevant information. The flow densities generated by each nozzle are listed in Table S1.

Water was introduced into the gas flow upstream of the Laval nozzle using a controlled evaporation mixing (CEM) system. A 1 litre reservoir maintained at 2 bar relative to atmospheric pressure connected to a liquid flow meter allowed flows of between 0.1 and 5 g hr$^{-1}$ of liquid water to be passed into an evaporation device heated to 353 K. A small fraction of the main carrier gas flow (Ar or $N_2$ for the $N_2$/$SF_6$ nozzle)) was also fed into the evaporation system through a mass flow controller, carrying water vapour into the reactor. To determine the gas-phase water concentration, the output of the CEM was passed into a 10 cm absorption cell at room temperature. The attenuation of the 185 nm line of a mercury pen-ray lamp was measured alternatively in the presence and absence of water vapour to yield values of the attenuated and non-attenuated intensities, I and $I_0$ respectively using a solar blind channel photomultiplier tube (CPM) with a spectral response in the 115 - 200 nm range operating in photon counting mode. In this way, only light from the 185 nm mercury line was detected. Typically, then number of photons falling on the CPM per second was recorded throughout the entire duration of the kinetic measurement. The mean number of photons per second was then taken to represent I. The same procedure was used in the absence of water vapour to obtain $I_0$ both before and after the kinetic measurement to assess the magnitude of any potential variations of the lamp intensity. The room temperature absorption cross-section of water vapour was taken from Sander et al.[26] to be 6.78 × $10^{-20}$ cm$^2$ so that its concentration could be calculated by the Beer-Lambert law. Under certain conditions of increasing high water flow, the measured water vapour absorption was seen to stabilize, indicating that the saturated vapour pressure limit of $H_2O$ in the cell for a given flow and total pressure had been reached. As a result, experiments were conducted below this threshold value. The output of the cell was connected to the reactor using a heating hose maintained at 353 K to avoid condensation. As the water vapour was diluted by at least a factor of five on entering the nozzle reservoir through mixing with the main carrier gas flow, we assume that no supplementary condensation losses occurred upstream of the Laval nozzle. In this way, gas-phase $H_2O$ concentrations as high as $4.5 \times 10^{14}$ molecule cm$^{-3}$ could be obtained in the cold flow. $CH(X^2\Pi_r)$ radicals were generated by the multiphoton dissociation of $CHBr_3$ molecules at 266 nm with ~ 25 mJ of pulse energy. The photolysis laser was aligned along the supersonic flow, to create a column of CH radicals of uniform density. $CHBr_3$ was entrained in the flow by bubbling a small flow of carrier gas through liquid $CHBr_3$ held at a known pressure.[27,28] An upper limit of $1 \times 10^{13}$ molecule cm$^{-3}$ was estimated for the gas-phase concentration of $CHBr_3$ in the experiments from its saturated vapour pressure.

$CH(X^2\Pi_r)$ radicals were followed through chemiluminescent emission from the electronically excited $OH(A^2\Sigma^+)$ (denoted OH*) product of the CH + $O_2$ reaction using a UV sensitive photomultiplier tube and an interference filter centered on the OH ($A^2\Sigma^+ \rightarrow X^2\Pi$) transition around 310 nm, in a similar manner to the measurements of Bergeat et al.[15] The $O_2$ concentration was constant for any single series of measurement and was kept to a minimum to avoid potential secondary reactions. The maximum $O_2$ concentration used during the experiments was $5 \times 10^{13}$ molecule cm$^{-3}$, but for most experiments the $O_2$ concentration was much lower. OH* detection allowed us to trace the entire CH temporal profile for each laser shot, drastically reducing the time required for signal acquisition. The time dependent chemiluminescence signals consisted of 500 time points acquired by a 500 MHz digital oscilloscope for a range of excess $H_2O$ concentrations. The signal was averaged over 768 photolysis laser cycles for each kinetic decay, the characteristic decay time for each trace depending on the H$_2$O concentration in the supersonic flow. The photolysis of CHBr$_3$ at 266 nm is likely to produce some excited state CH(a$^4\Sigma$) radicals in our experiments which might interfere with our investigation of the kinetics of the CH(X$^2\Pi_r$) + H$_2$O reaction. It has been estimated[29] that the photolysis of CHBr$_3$ at 248 nm produces five times more CH radicals in the X$^2\Pi_r$ state than in the a$^4\Sigma$ state. Consequently we expect that there will be a slightly smaller relative yield of CH(a$^4\Sigma$) at 266 nm given the lower energy available. Indeed OH* is produced by the CH(a$^4\Sigma$) + O$_2$ reaction with a room temperature rate constant of 2.6 × 10$^{-11}$ cm$^3$ molecule$^{-1}$ s$^{-1}$.[30] Fortunately, CH(a$^4\Sigma$) radicals have been shown to react slowly with H$_2$O with a room temperature rate constant of less than 2 × 10$^{-13}$ cm$^3$ molecule$^{-1}$ s$^{-1}$.[31]

The intercepts of second-order plots such as those shown in Figure 2 and in Figures S1-S5 represent the sum of several contributions. As the chemiluminescent OH*emission of the CH(X$^2\Pi_r$) + O$_2$ reaction was used to follow CH(X$^2\Pi_r$), a small fraction of these radicals are consumed. The temperature dependent rate constants for the CH + O$_2$ reaction have been measured previously[32] allowing us to estimate a maximum pseudo-first-order loss rate for CH of ~ 3000 s$^{-1}$ at 50 K. Test experiments performed at 296 K with different fixed O$_2$ concentrations each over a range of H$_2$O concentrations yielded second-order plots with the same slope. The only noticeable difference was a change in the intercept values, indicating that the products of this tracer reaction did not interfere with CH + H$_2$O kinetics. Moreover, a small part of the OH* emission could arise from the CH(a$^4\Sigma$) + O$_2$ reaction. Nevertheless, this will have a constant contribution to the intercept for any series of measurements as CH(a$^4\Sigma$) reacts only slowly with water. CH(X$^2\Pi_r$) radicals were also lost through their rapid reaction with the precursor molecule CHBr$_3$. Nevertheless, test experiments performed at 296 K showed that the slopes of second-order plots did not change for a range of initial CHBr$_3$ concentrations, contributing only to a change in the intercept value. Finally, diffusional loss of CH radicals from the area illuminated by the probe laser, might also have contributed to the intercept.

**Supplementary Figures:**

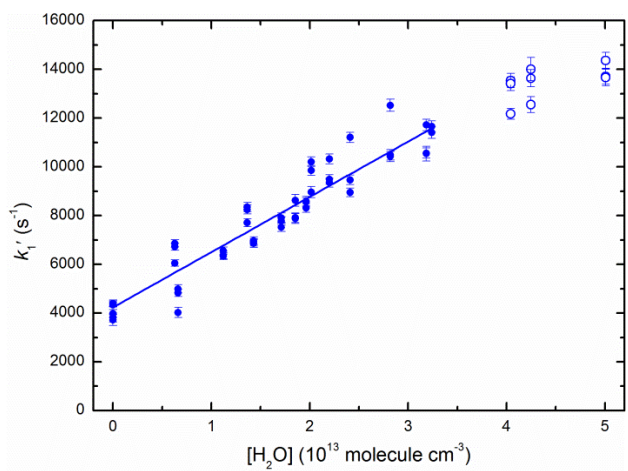

**Figure S1.** (●) Plots of the $k_1'$ values at 75 K as a function of the $H_2O$ concentration. (○) Datapoints obtained at 75 K with $H_2O$ concentrations above the threshold for cluster formation. The uncertainties are obtained from non-linear fits to decays similar to those presented in Figure 1. The solid lines represent fits using a $1/\sigma^2$ weighting function. $\sigma$ is the statistical uncertainty of individual $k_1'$ values.

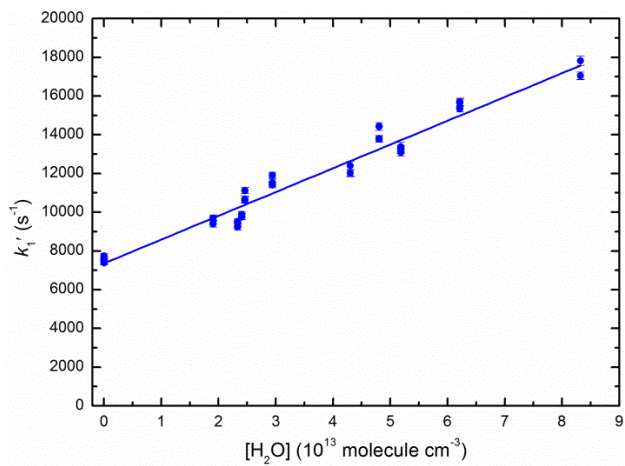

**Figure S2.** (●) Plots of the $k_1'$ values at 127 K as a function of the $H_2O$ concentration. The uncertainties are obtained from non-linear fits to decays similar to those presented in Figure 1. The solid lines represent fits using a $1/\sigma^2$ weighting function. $\sigma$ is the statistical uncertainty of individual $k_1'$ values.

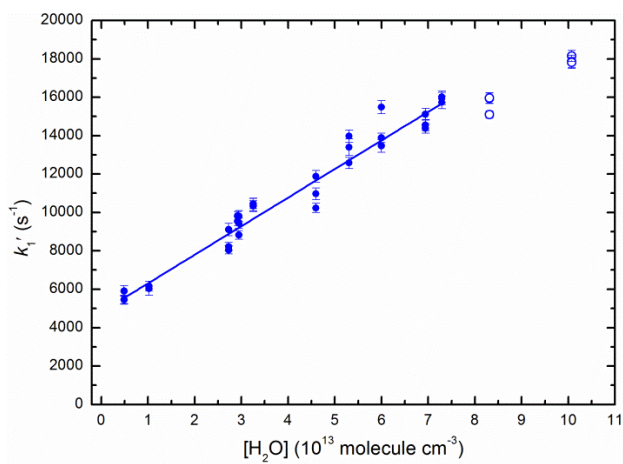

**Figure S3.** (●) Plots of the $k_1'$ values at 134 K as a function of the $H_2O$ concentration. (○) Datapoints obtained at 134 K with $H_2O$ concentrations above the threshold for cluster formation. The uncertainties are obtained from non-linear fits to decays similar to those presented in Figure 1. The solid lines represent fits using a $1/\sigma^2$ weighting function. $\sigma$ is the statistical uncertainty of individual $k_1'$ values.

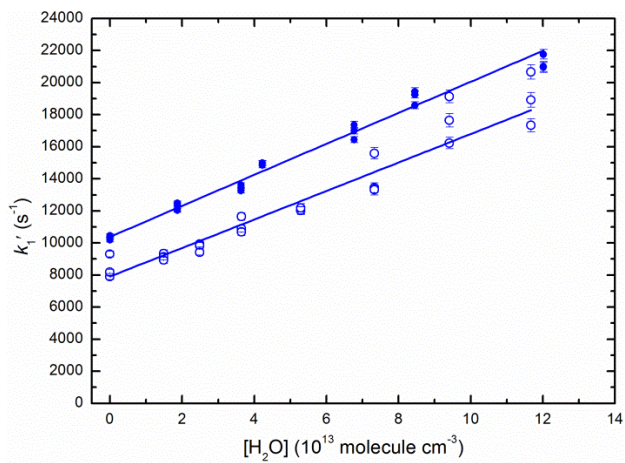

**Figure S4.** Plots of the $k_1'$ values at 241 K as a function of the H$_2$O concentration(●, ○). The two datasets were obtained with different precursor molecule concentrations to check for secondary reactions. The uncertainties are obtained from non-linear fits to decays similar to those presented in Figure 1. The solid lines represent fits using a $1/\sigma^2$ weighting function. $\sigma$ is the statistical uncertainty of individual $k_1'$ values.

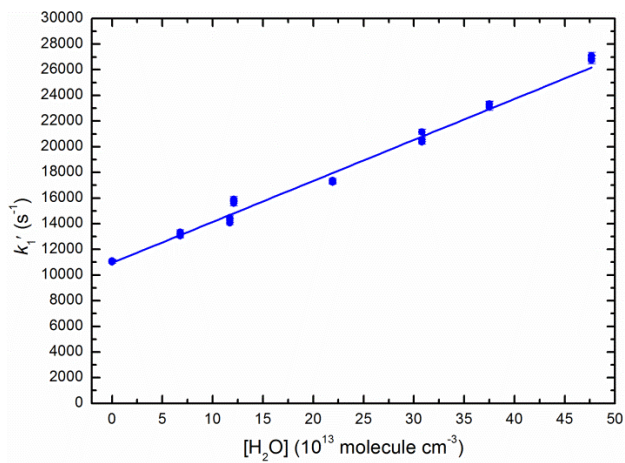

**Figure S5.** (●) Plots of the $k_1'$ values at 296 K as a function of the H$_2$O concentration using a mixture of N$_2$ / SF$_6$ as the carrier gas. The uncertainties are obtained from non-linear fits to decays similar to those presented in Figure 1. The solid lines represent fits using a $1/\sigma^2$ weighting function. $\sigma$ is the statistical uncertainty of individual $k_1'$ values.

**Supplementary Tables:**

**Table S1.** Second-order rate constants for the CH + $H_2O$ reaction

| T (K) | [$H_2O$] (10$^{13}$ molecule cm$^{-3}$) | Flow Density (10$^{16}$ molecule cm$^{-3}$) | Carrier Gas | $k_1$ (10$^{-11}$ cm$^3$ molecule$^{-1}$ s$^{-1}$) |
|---|---|---|---|---|
| 296 | 0 – 44.7 | 19.2 | $N_2$/$SF_6$ | 3.20 ± 0.33[a] |
| 241 | 0 – 12.0 | 9.0 ± 0.2[b] | $N_2$/$SF_6$ | 9.28 ± 1.1 |
| 296 | 0 – 22.7 | 5.2 – 63.7 | Ar | 3.57 ± 0.37[a] |
| 134 | 0.5 – 7.3 | 45.4 ± 1.2 | Ar | 14.9 ± 1.6 |
| 127 | 0 – 8.3 | 12.6 ± 0.3 | Ar | 12.3 ± 1.3 |
| 75 | 0 – 5.0 | 14.7 ± 0.6 | Ar | 22.7 ± 2.5 |
| 50 | 0 – 3.3 | 25.9 ± 0.9 | Ar | 23.0 ± 2.5 |

[a]The rate constant errors are the combined statistical and systematic uncertainties. Statistical uncertainties are included at the level of a single standard deviation. The systematic uncertainties, resulting from potential errors in the measured flow velocities, densities, temperatures and flow rates, are estimated at 10%. The corresponding standard deviation was estimated by using a uniform probability distribution assuming equal probability for the value to lie within 10% of the mean value. [b]The errors on the flow densities are calculated from separate measurements of the impact pressure using a Pitot tube as a function of distance from the Laval nozzle and the stagnation pressure within the reservoir.

**Table S2.** Previously measured second-order rate constants for the CH + $H_2O$ reaction

| T (K) | Pressure (Torr) | [$H_2O$] (10$^{14}$ molecule cm$^{-3}$) | Carrier Gas | $k_1$ (10$^{-11}$ cm$^3$ molecule$^{-1}$ s$^{-1}$) | |
|---|---|---|---|---|---|
| 298 | 11-205 | | | 4.48 ± 0.85 | Bosnali and Perner[19] |
| 291 | 70 | 2.2 – 13.6 | He | 1.77 ± 0.11 | Blitz et al.[7] |
| 293 | 200 | 7.1 – 45.6 | He | 1.27 ± 0.19 | |
| 373 | | 1.8 – 11.5 | He | 1.01 ± 0.08 | |
| 473 | | 1.4 – 8.9 | He | 0.81 ± 0.08 | |
| 573 | | 1.8 – 11.6 | He | 0.71 ± 0.08 | |

| | | | | |
|---|---|---|---|---|
| 673 | 1.8 – 11.2 | He | 0.47 ± 0.05 | |
| 298 | 20 | Ar | 3.18 ± 0.09 | Zabarnick et al.[8] |
| 298 | 100 | Ar | 3.42 ± 0.12 | |
| 297 | 300 | Ar | 3.08 ± 0.09 | |
| 371 | 100 | Ar | 2.53 ± 0.05 | |
| 440 | 100 | Ar | 2.31 ± 0.04 | |
| 508 | 100 | Ar | 1.97 ± 0.06 | |
| 584 | 100 | Ar | 1.87 ± 0.04 | |
| 669 | 100 | Ar | 1.63 ± 0.06 | |
| 330 | 1.4 | He | 0.90 ± 0.40 | Bergeat et al.[15] |

**Table S3.** Continuous supersonic flow characteristics

| Laval nozzle | Mach2 $N_2/SF_6$ | Mach2$N_2$ (used with Ar) | Mach2Ar | Mach3Ar | Mach4Ar |
|---|---|---|---|---|---|
| Mach number | 2.1 ± 0.01[a] | 2.0 ± 0.03 | 1.9 ± 0.03 | 3.0 ± 0.1 | 3.9 ± 0.1 |
| Carrier gas | $(N_2/SF_6)$[b] | Ar | Ar | Ar | Ar |
| Impact pressure (Torr) | 11.5 | 10.5 | 36.3 | 15.3 | 29.6 |
| Stagnation pressure (Torr) | 18.6 | 13.9 | 45.7 | 34.9 | 113 |
| Temperature (K) | 241 ± 1[a] | 127 ± 2 | 134 ± 2 | 75 ± 2 | 50 ± 1 |
| Mean flow velocity (ms$^{-1}$) | 282 ± 2[a] | 419 ± 3 | 409 ± 3 | 479 ± 2 | 505 ± 1 |

[a]The errors on the Mach number, temperature and mean flow velocity are calculated from separate measurements of the impact pressure using a Pitot tube as a function of distance from the Laval nozzle and the stagnation pressure within the reservoir. [b]Mole fraction $N_2$ / $SF_6$ = 0.4 / 0.6.